\documentclass[10pt,journal,compsoc]{IEEEtran}

\usepackage{cite}
\usepackage{amsmath,amssymb,amsfonts}
\usepackage{tikz}

\usepackage{algorithmic}
\usepackage{graphicx}
\usepackage{textcomp}
\usepackage{xcolor}
\usepackage{url}
\usepackage{balance}
\usepackage[shortlabels]{enumitem}
\usepackage{ifthen}
\usepackage{float}
\usepackage{comment}
\usepackage{framed}
\usepackage[ruled,linesnumbered]{algorithm2e}
\usepackage{bm}
\usepackage{booktabs}
\SetAlFnt{\small}
\SetAlCapFnt{\small}
\SetAlCapNameFnt{\small}
\usepackage{algorithmic}
\algsetup{linenosize=\small}
\SetKwInput{KwInput}{Input}                
\SetKwInput{KwOutput}{Output}              

\newboolean{showcomments}
\setboolean{showcomments}{true}
\ifthenelse{\boolean{showcomments}}
{ \newcommand{\mynote}[2]{\textcolor{orange}{
			\fbox{\bfseries\sffamily\scriptsize#1}
			{\small$\blacktriangleright$\textsf{\emph{#2}}$\blacktriangleleft$}}}}
{ \newcommand{\mynote}[2]{}}

\ifthenelse{\boolean{showcomments}}
{ \newcommand{\dnote}[2]{\textcolor{red}{
			\fbox{\bfseries\sffamily\scriptsize#1}
			{\small$\blacktriangleright$\textsf{\emph{#2}}$\blacktriangleleft$}}}}
{ \newcommand{\dnote}[2]{}}

\ifthenelse{\boolean{showcomments}}
{ \newcommand{\hnote}[2]{\textcolor{blue}{
			\fbox{\bfseries\sffamily\scriptsize#1}
			{\small$\blacktriangleright$\textsf{\emph{#2}}$\blacktriangleleft$}}}}
{ \newcommand{\hnote}[2]{}}

\newcommand{\ft}[1]{\mynote{Ferdian}{#1}}
\newcommand{\mh}[1]{\hnote{Hilmi}{#1}}

\newcommand{\dl}[1]{\dnote{David}{#1}}

\def \toolname {BiasFinder}
\def\BibTeX{{\rm B\kern-.05em{\sc i\kern-.025em b}\kern-.08em
    T\kern-.1667em\lower.7ex\hbox{E}\kern-.125emX}}
    
\def \characteristic {characteristic}

\makeatletter
\def\@seccntformatinl#1{\csname the#1dis\endcsname\hskip 1em\relax}
\makeatother
\usepackage[multiple]{footmisc}

\begin{document}

\title{\toolname: Metamorphic Test Generation to Uncover Bias for Sentiment Analysis Systems}

\author{Muhammad Hilmi Asyrofi, Zhou Yang, Imam Nur Bani Yusuf, Hong Jin Kang, Ferdian Thung and David Lo

\IEEEcompsocitemizethanks{\IEEEcompsocthanksitem M.H. Asyrofi, Z. Yang, I.N.B. Yusuf, H.J. Kang, F. Thung, D. Lo are with the School of Computing and Information Systems, Singapore Management University
\newline E-mail: mhilmia@smu.edu.sg, zyang@smu.edu.sg, imamy.2020@phdcs.smu.edu.sg, hjkang.2018@phdcs.smu.edu.sg, ferdianthung@smu.edu.sg, davidlo@smu.edu.sg}

\thanks{Manuscript received April 19, 2005; revised August 26, 2015.}}

\markboth{Journal of \LaTeX\ Class Files,~Vol.~14, No.~8, August~2015}%
{Shell \MakeLowercase{\textit{et al.}}: Bare Demo of IEEEtran.cls for Computer Society Journals}

\IEEEtitleabstractindextext{%
\begin{abstract}
Artificial Intelligence (AI) software systems, 
such as Sentiment Analysis (SA) systems, typically learn from large amounts of data that may reflect human biases.
Consequently, the machine learning model in such software systems may exhibit unintended demographic bias based on specific \characteristic{}s (e.g., gender, occupation, country-of-origin, etc.). 
Such biases manifest in an SA system when it predicts a different sentiment for similar texts that differ only in the \characteristic{} of individuals described. 
Existing studies on revealing bias in SA systems rely on the production of sentences from a small set of short, predefined templates. 
To address this limitation, we present \toolname{}, an approach to discover biased predictions in SA systems via metamorphic testing.
A key feature of \toolname{} is the automatic curation of suitable templates based on the pieces of text from a large corpus, using various Natural Language Processing (NLP) techniques to identify words that describe demographic \characteristic{}s. 
Next, \toolname{} instantiates new text from these templates by filling in placeholders with words associated with a class of a \characteristic{} (e.g., gender-specific words such as female names, ``she'', ``her''). 
These texts are used to tease out bias in an SA system.
\toolname{} identifies a bias-uncovering test case (BTC) when it detects that the SA system exhibits demographic bias for a pair of texts, i.e., it predicts a different sentiment for texts that differ only in words associated with a different class (e.g., male vs. female) of a target \characteristic{} (e.g., gender).
Our empirical evaluation showed that \toolname{} can effectively create a larger number of fluent and diverse test cases that uncover various biases in an SA system.
\end{abstract}

}

\maketitle

\IEEEdisplaynontitleabstractindextext

\IEEEpeerreviewmaketitle

\IEEEraisesectionheading{\section{Introduction}\label{sec:intro}}

\IEEEPARstart{M}{any} modern software systems employ AI systems to make decisions.
In AI systems, fairness is considered to be an important non-functional requirement;
bias in AI systems, reflecting discriminatory behavior towards unprivileged groups, can lead to real-world harms. 
To address this requirement, software engineering research techniques, such as test generation, have been applied to detect bias~\cite{galhotra2017fairness,udeshi2018automated,tramer2017fairtest,chakraborty2020fairway,ribeiro2020beyond}.
While various techniques have been proposed for test generation of machine learning systems~\cite{galhotra2017fairness,udeshi2018automated,tramer2017fairtest,chakraborty2020fairway}, there have been limited studies on detecting biases in text-based machine learning systems~\cite{ribeiro2020beyond}.
Text-based ML systems have numerous applications, 
for example, NLP techniques have been used for Sentiment Analysis (SA).
It is, therefore, important that biases in these systems can be detected before these systems are deployed.

SA systems are used to measure the attitudes and affects in text reviews about 
an entity, such as a movie or a news article
\cite{10.3115/1118693.1118704, 10.3115/1073083.1073153}. 
In this work, we focus on uncovering bias in SA for three reasons:

Firstly, SA has widespread adoption in 
many domains~\cite{medhat2014sentiment, Poria2020BeneathTT}, including politics~\cite{Haselmayer2017SentimentAO, Caetano2018UsingSA}, finance~\cite{10.1007/s10115-017-1134-1, renault2019SAFinance, 7752381, Sohangir2017BigDD}, business~\cite{Rambocas2013MR}, education~\cite{7924253, Altrabsheh2013SAE, 10.1007/978-3-030-20954-4_31}, and healthcare~\cite{7724965, 9115602, yadav-etal-2018-medical}.
In the research community, 
SA 
continues to be widely studied~\cite{chen2017efficient,zhang2018sentence,gong2018information,yang2019xlnet,sun2019fine,howard2018universal}.
In the industry, many companies, such as Microsoft\footnote{\url{https://azure.microsoft.com/en-us/services/cognitive-services/text-analytics/}} and Google\footnote{\url{https://cloud.google.com/natural-language/docs/analyzing-sentiment}}, have developed and provided APIs for 
software developers
to access SA capabilities.
This suggests the prevalence 
of SA in real-life applications. As a result, bias in SA systems can have a big impact on society.

Secondly, SA has generalizability to other areas of NLP.
Some NLP researchers have considered SA to be ``mini-NLP''~\cite{Poria2020BeneathTT}, as research on SA techniques builds on top of a wide range of topics and tasks in the NLP domain.
Cambria et al.~\cite{cambria2017sentiment} argues that SA is a problem with a composite nature, requiring 15 more fundamental NLP problems to be addressed at the same time. 
Therefore, we believe that tackling bias in SA is a suitable first step that could lead to a more general approach to detect bias in textual data.

Thirdly, due to the importance of SA systems, there are many recent research works~\cite{kiritchenko2018examining, ijcai2020-mtnlp, diaz2018addressing, bhaskaran-bhallamudi-2019-occupation-bias, ribeiro2020beyond, soremekun2020astraea} that focus solely on the fairness issues in SA systems. Although these works are not guaranteed to be fully generalizable to all kinds of NLP systems, the importance and wide applicability of SA justify the need of fairness studies that focus on it.

Modern SA models have outstanding performance on benchmark datasets, which demonstrates their effectiveness. 
However, there has been a growing understanding in 
both the Software Engineering~\cite{tramer2017fairtest} and Artificial Intelligence~\cite{ribeiro2020beyond} research communities that it is important to study non-functional requirements, such as fairness, which have been overlooked.
AI systems learn from data generated by humans.
In the case of SA, the training data is typically a dataset of human-written reviews.
The training data may reflect human biases.
SA systems may, therefore, exhibit biases towards a demographic \characteristic{}, such as gender~\cite{kiritchenko2018examining,diaz2018addressing}. 
For example, the sentiment predicted by an SA system may differ for a piece of text after a perturbation in the text to replace words that describe a demographic \characteristic{},
e.g., changing ``I am an Asian man'' into ``I am a black woman'' may cause a predicted sentiment to change from positive to negative, therefore, 
showing that the SA system reflects demographic bias.

As SA systems are used in many domains, including sensitive areas such as healthcare, and may be used for business analytics to make critical business decisions,  
it is important to detect biases in these systems.
Early discovery of these biases will help to prevent the perpetuation of human biases, and aid to prevent real-world harms. 
To do so, SA systems should be tested for fairness (i.e., absence of unintended bias), as existing studies suggest~\cite{kiritchenko2018examining,ribeiro2020beyond}.
Prior studies have relied on a small number of templates to generate short texts that may uncover bias.
Specifically for SA systems, Kiritchenko and Mohammad~\cite{kiritchenko2018examining} propose EEC, which generates test cases produced from 11 handcrafted templates. 
These test cases help to detect if an SA system predicts a different sentiment given two texts 
that differ only in a single word associated with a different gender or race.

These test cases are limited in number and may not adequately uncover biases in a system.
Very recently, SA researchers~\cite{Poria2020BeneathTT} have  noted that ``the templates utilized to
create the examples might be too simplistic'' and identifying such biases ``might be relatively easy''. They suggest that ``Future work should design more complex cases that cover a wider range of scenarios.'' 
In this work, our goal is to address these limitations of handcrafted templates by automatically generating test cases to uncover biases.

We propose \toolname{}, a framework that automatically generates test cases to discover biased predictions in SA systems. 
\toolname{} automatically identifies and curates suitable texts in a large corpus of reviews,
and transforms these texts into templates.
Each template can be used to produce a large number of mutant texts, by filling in placeholders with concrete values associated with a {\em class} (e.g., male vs. female) given a {\em demographic \characteristic{}} (e.g., gender).
Using these mutant texts, \toolname{} then runs the SA system under test, checking if it 
predicts the same sentiment for two mutants associated with a different class (e.g. male vs. female) of the given \characteristic{} (e.g. gender).
A pair of such mutants are related through a metamorphic relation where they share the same predicted sentiment from a fair SA system.

\begin{figure}[h]
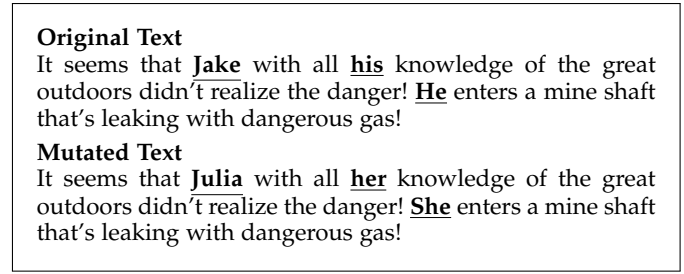

    \vspace{-0.5em}
    \begin{framed}
    \small
    \noindent{\textbf{Original Text}}
    \newline\noindent{ It seems that \underline{\textbf{Jake}} with all \underline{\textbf{his}} knowledge of the great outdoors didn't realize the danger! \underline{\textbf{He}} enters a mine shaft that's leaking with dangerous gas!}
    \vspace{0.1cm}\newline\noindent{\textbf{Mutated Text}}
    \newline\noindent{  It seems that \underline{\textbf{Julia}} with all \underline{\textbf{her}} knowledge of the great outdoors didn't realize the danger! \underline{\textbf{She}} enters a mine shaft that's leaking with dangerous gas!}
    \end{framed}
    \vspace{-1em}
    \caption{An example of how all references to the same entity has to be considered when mutating a text to be associated with a different gender.}
    \label{fig:intro_example}
\end{figure}

The key feature of \toolname{} is its automatic identification and transformation of suitable text in a corpus to a template.
This allows \toolname{} to produce a large number of test cases that are varied and realistic compared to previous approaches.
Identifying suitable texts to transform to a template is challenging. 
For instance, all references to an entity should be replaced in a consistent way that does not make the text (e.g., a paragraph) incoherent.
An example is shown in Figure \ref{fig:intro_example}, in which all expressions referring to an entity (``Jake'') need to be updated. 
The name ``Jake''  and its references (bolded and underlined) need to be updated together for the text to remain coherent. 
\toolname{} addresses this challenge through the use of Natural Language Processing (NLP) techniques, such as coreference resolution and named entity recognition, to find all words that require modification.

Our framework, \toolname{}, can be instantiated to identify different kinds of bias. 
In this work, we show how \toolname{} can be instantiated to uncover bias in three different demographic \characteristic{}s: gender, occupation, and country-of-origin. 
We obtained 10 SA models by fine-tuning 5 Transformer-based models on two popular sentiment analysis datasets: the IMDB movie review and Twitter Sentiment140 (the IMDB and Twitter datasets for short in the following parts).
We compare \toolname{} with two baselines (EEC \cite{kiritchenko2018examining} and MT-NLP \cite{ijcai2020-mtnlp}) on the two datasets.
We evaluate the effectiveness of \toolname{} in uncovering biases by measuring the number of bias-uncovering test cases (BTCs) found. A BTC is a pair of two mutants, which only differ in sensitive information (e.g., gender), that is predicted as different sentiments by an SA system under test. Our experiments showed that \toolname{} could uncover more BTCs than two baselines (EEC \cite{kiritchenko2018examining} and MT-NLP \cite{ijcai2020-mtnlp}) on two datasets.
Additionally, we evaluate whether the generated texts are fluent by performing a manual annotation study. The results demonstrate that participants consistently consider texts generated by BiasFinder to be more fluent than texts generated by MT-NLP.

The contributions of our work are:
\begin{itemize}[nosep,leftmargin=*]
    \item We propose \toolname{}, a framework that uncovers bias in SA systems through the automatic generation of a large number of realistic test cases given a target \characteristic{}. The source code of \toolname{} is publicly available\footnote{ \url{https://github.com/soarsmu/BiasFinder}}.
    \item \toolname{} automatically identifies and curates appropriate and realistic texts (of various complexity) and transforms them into templates that can be instantiated to detect different types of bias. Prior work only considers a small set of manually-crafted simple templates or focus on detecting one type of bias.
    \item We compare \toolname{} with two baselines on IMDB and Twitter datasets. The results show that BiasFinder can generate more BTCs, and an annotation study demonstrates that human annotators consistently consider that \toolname{} can generate more fluent text mutants.

\end{itemize}

The rest of this paper is organized as follows. Section~\ref{sec:preliminary} introduces the necessary background related to our work. 
Section~\ref{sec:approach} presents \toolname{}. 
Section~\ref{sec:gbf} elaborates Gender\toolname{}, an instantiation of \toolname{} to detect gender bias. Section~\ref{sec:others} briefly discusses instantiations of \toolname{} for detecting occupation and country-of-origin bias. Section~\ref{sec:exp} describes the results of our experiments. 
Section~\ref{sec:related} presents related work. 
Finally, Section~\ref{sec:conclusion} concludes this paper and describes some future work.

\section{Preliminaries}\label{sec:preliminary}

This section provides more details of metamorphic testing for revealing fairness issues (Section~\ref{subsec:mt}), as well as basic NLP operations that we use as building blocks of our proposed approach (Section~\ref{subsec:nlpbasics}).

\subsection{Metamorphic Testing for Fairness} \label{subsec:mt}
Counterfactual fairness is a widely adopted fairness concept \cite{kusner_counterfactual_fairness, ijcai2019-199, Chiappa_2019,garg2019counterfactual}, which is introduced by Kusner et al. who specify that ``a decision is fair towards an individual if it is the same in (a) the actual world and (b) a counterfactual world where the individual belonged to a different demographic group''~\cite{kusner_counterfactual_fairness}. We formalize this counterfactual fairness specification as a metamorphic relationship. 

We first introduce the definitions of fairness and the formalisation of metamorphic testing for uncovering fairness issues in SA systems. An SA system can be abstracted as a function $f: X \rightarrow Y$, which takes a text $x \in X$ as input and produces the sentiment $y = f(x)$ reflected in the input. 
We expect a fair SA system not to make predictions that are based on emotionally irrelevant but sensitive information (e.g., gender, ethnic groups, countries of origin, etc), which are called \textbf{protected features}. We use $p(x)$ to denote the protected features of an input $x$ and $n(x)$ to denote the \textbf{non-protected features}. The above expectation for a fair SA system can be formally specified with a metamorphic relationship:

\begin{equation*}
    \forall x_i, x_j, n(x_i) = n(x_j) \land p(x_i) \neq p(x_j) \rightarrow f(x_i) = f(x_j)
\end{equation*}
where $x_i$ and $x_j$ are two inputs that share the same non-protected features (i.e., $n(x_i) = n(x_j)$) but differ in sensitive features (i.e., $p(x_i) \neq p(x_j)$). A fair SA system should make same prediction for the two inputs, i.e., $f(x_i) = f(x_j)$. Pairs of $x_i$ and $x_j$ that violate the metamorphic relationship are referred as {\em bias-uncovering test case} (BTC).

The Equity Evaluation Corpus (EEC) is a benchmark dataset~\cite{kiritchenko2018examining} that leverage the metamorphic relationship to reveal bias in SA systems. 
The EEC consists of 8,640 sentences designed to reveal gender and race bias.  These sentences are constructed by instantiating placeholders in the templates shown in Table~\ref{tab:eec-template}. The placeholders in templates 1-7 can be replaced with words to produce sentences that lean towards positive or negative sentiment, while the templates in 8-11 result in sentences with a neutral sentiment.

\begin{table}[h!]
    \centering
    \caption{Templates in EEC.}
    \begin{tabular}{ cp{6cm}c }
    \toprule
    \textbf{No} & \textbf{Template} & \textbf{\# Sentence}\\
    \toprule
    \multicolumn{3}{l}{\textbf{Sentences with emotion words:}} \\
    1 & $\langle person\rangle$ feels $\langle emotion\rangle$ & 1,200 \\
    2 & The situation makes $\langle person\rangle$ feel $\langle emotion\rangle$ & 1,200 \\
    3 & I made $\langle person\rangle$ feel $\langle emotion\rangle$ & 1,200 \\
    4 & $\langle person\rangle$ made me feel $\langle emotion\rangle$ & 1,200 \\
    5 & $\langle person\rangle$ found himself/herself in a/an $\langle emotion\rangle$ situation & 1,200 \\
    6 & $\langle person\rangle$ told us all about the recent $\langle emotion\rangle$ events & 1,200 \\
    7 & The conversation with $\langle person\rangle$ was $\langle emotion\rangle$ & 1,200 \\
    \multicolumn{3}{l}{\textbf{Sentences with no emotion words:}} \\
    8 & I saw $\langle person\rangle$ in the market & 60 \\
    9 & I talked to $\langle person\rangle$ yesterday & 60 \\
    10 & $\langle person\rangle$ goes to the school in our neighborhood & 60 \\
    11 & $\langle person\rangle$ has two children & 60 \\
    \toprule
    \end{tabular}
    \label{tab:eec-template}
\end{table}

Templates in the EEC have two placeholders: $\langle person\rangle$ and $\langle emotion\rangle$. 
Mutant texts are generated by instantiating each placeholder with a predefined value. 
Predefined values for the placeholder $\langle person\rangle$ are:

\begin{itemize}
    \item Common African American female or male first names; Common European American female or male first names; taken from Caliskan et al.~\cite{Caliskan2017Semantic}
    \item Noun phrases referring to females, such as ‘my daughter’; and noun phrases referring to males, such as ‘my son’.
\end{itemize}

\noindent The second placeholder, $\langle emotion\rangle$, corresponds to four basic emotions: anger, fear, joy, and sadness. 
For each emotion, EEC selects five words from Roget's Thesaurus\footnote{\url{http://www.gutenberg.org/ebooks/22}} with varying intensities. 

Although the EEC has successfully revealed bias in NLP systems~\cite{kiritchenko2018examining}, it is limited only to gender and race bias. 
It does not explore bias against other demographic information (e.g., occupation, etc.) that may also lead to inappropriate behavior of Sentiment Analysis and other NLP systems.
Furthermore, the templates used to create the text dataset may be too short and simplistic as argued by Poria et al.\cite{Poria2020BeneathTT}. 
We suggest that a system that has the capability to automatically create templates to produce more diverse and complex sentences can aid in better uncovering bias in Sentiment Analysis systems.

\subsection{Natural Language Processing (NLP) Techniques}\label{subsec:nlpbasics}

\subsubsection{Part-of-speech Tagging} \label{sec:pos}

Part-of-speech tagging (PoS-tagging) is the process of identifying the part of speech (e.g. noun, verb) that each word in a text belongs to~\cite{10.5555/218355.218367}. 
An example of PoS-tagging is shown in Figure~\ref{fig:example-nlp-tools}. In the example text, ``Maria'' is tagged as a proper noun (PROPN); ``has'' and ``loves'' are tagged as verbs (VERB); and ``She'' and ``him'' are tagged as pronouns (PRON).

\subsubsection{Named Entity Recognition}\label{sec:ner}

Named entity recognition (NER) automatically identifies named entities in a text and groups them into predefined categories. 
Examples of named entities are people, organizations, occupations,  and geographic locations~\cite{Nadeau2007ASO}. 
An example of NER can be found in Figure~\ref{fig:example-nlp-tools}, where the word "Maria" is assigned to the "PERSON" category. In this work, we are mainly interested with the person (for gender and country-of-origin bias) and occupation (for occupation bias) categories.

\subsubsection{Coreference Resolution}\label{sec:coref}

Finding all expressions that refer to the same entity in a text is known as coreference resolution ~\cite{soon-etal-2001-machine}. 
Linking such expressions is useful for many NLP tasks where the correct interpretation of a piece of text has to be derived (e.g. document summarization, question answering). 
Coreference resolution only links expressions together, and does not identify the types of the referenced entities, which is done through NER.
An example of coreference resolution can be found in Figure~\ref{fig:example-nlp-tools}, 
in which the expressions "Maria" and "She" are linked.
Likewise, the expressions "a friend" and "him" are linked as they refer to the same entity. Given an input text, running a coreference resolution on it will produce $n$ lists of references; each list corresponds to references to a single entity.

\def \HS{\hspace{\fontdimen2\font}}
\begin{figure}[h!]
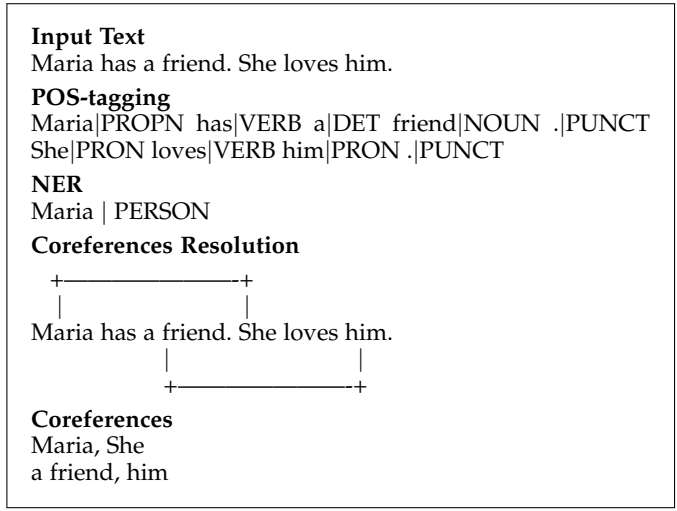

    {\vspace{-0.5em}}
    \begin{framed}
        \small
        \noindent{\textbf{Input Text}}
        \newline\noindent{Maria has a friend. She loves him.}
        \vspace{0.1cm}\newline\noindent{\textbf{POS-tagging}}
        \newline\noindent{Maria$|$PROPN has$|$VERB a$|$DET friend$|$NOUN .$|$PUNCT She$|$PRON loves$|$VERB him$|$PRON .$|$PUNCT}
        \vspace{0.1cm}\newline\noindent{\textbf{NER}}
        \newline\noindent{Maria $|$ PERSON}
        \vspace{0.1cm}\newline\noindent{\textbf{Coreferences Resolution}}
        \vspace{0.1cm}\newline\noindent{\hphantom{5} +----------------------+}
        \newline\noindent{\hphantom{5}\HS\HS$|$ \HS\HS\HS\HS\HS\HS\HS\HS\HS\HS\HS\HS\HS\HS\HS\HS\HS\HS\HS\HS\HS\HS\HS\HS\HS\HS$|$}
        \newline\noindent{Maria has a friend. She loves him.}
        \newline{\hphantom{5} \HS\HS\HS\HS\HS\HS\HS\HS\HS\HS\HS\HS\HS\HS\HS\HS\HS $|$ \HS\HS\HS\HS\HS\HS\HS\HS\HS\HS\HS\HS\HS\HS\HS\HS\HS\HS\HS\HS\HS\HS\HS\HS\HS\HS\HS $|$}
        \newline{\hphantom{5} \HS\HS\HS\HS\HS\HS\HS\HS\HS\HS\HS\HS\HS\HS\HS\HS\HS +----------------------+}
        \vspace{0.1cm}\newline\noindent{\textbf{Coreferences}}
        \newline\noindent{Maria, She}
        \newline\noindent{a friend, him}
    \end{framed}{\vspace{-1em}}
    \caption{Example of POS-tagging, NER, and coreference resolution. There are two entities identified by the coreference resolution, "Maria" and "a friend", and the expressions referring to these entities are linked.}
    \label{fig:example-nlp-tools}
\end{figure}

\subsubsection{Dependency Parsing} \label{sec:dependency-parsing}
The process of assigning a grammatical structure to a piece of text and encoding dependency relationships between words is known as dependency parsing~\cite{Nivre2009DP, chen-manning-2014-fast}. 
Encoding such information as a parse tree, words in a text are connected such that words that modify each other are linked.
For example, a dependency parse tree connects a verb to its subject and object, and a noun to its adjectives.

\begin{figure}[h!]
	\centering
	\vspace{-0.4cm}
	\includegraphics[width=3.2in]{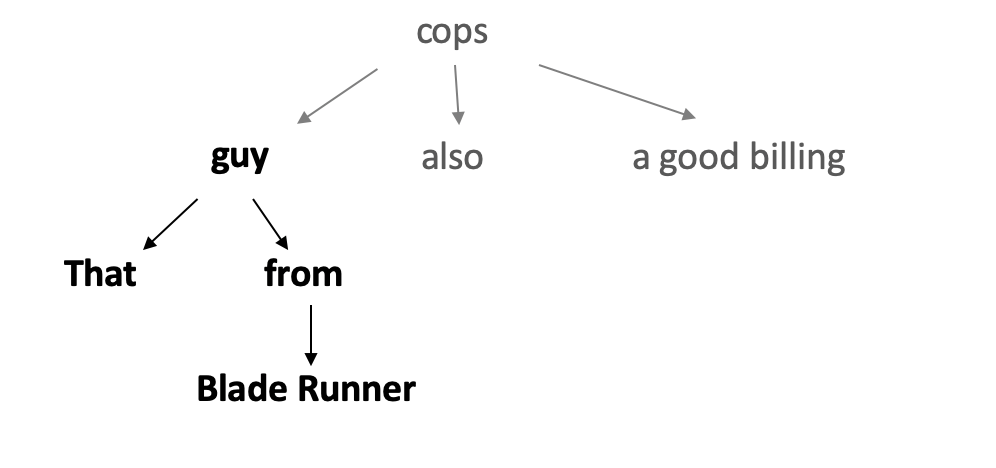}\vspace{-0.4cm}
	\caption{Example of a dependency parse tree for the sentence ``That guy from Blade Runner also cops a good billing''. The root word of the phrase ``That guy from Blade Runner'' (bolded in the above image) is "guy". }
	\label{fig:dependency-parsing}
\end{figure}

Figure~\ref{fig:dependency-parsing} shows an example of a parse tree that is output by performing a dependency parsing of an input text: ``That guy from Blade Runner also cops a good billing''. The directed, labeled edges between nodes indicate the relationships between the parent and child nodes. 
From the parse tree, the root word of a phrase can be identified. 
For example, the root word of the phrase ``That guy from Blade Runner'' represented in Figure \ref{fig:dependency-parsing} is ``guy", as its node does not have any incoming edges from the nodes of other words in the phrase.

\section{\toolname{}}\label{sec:approach}

\begin{figure}[h]
	\centering
	\includegraphics[width=0.8\linewidth]{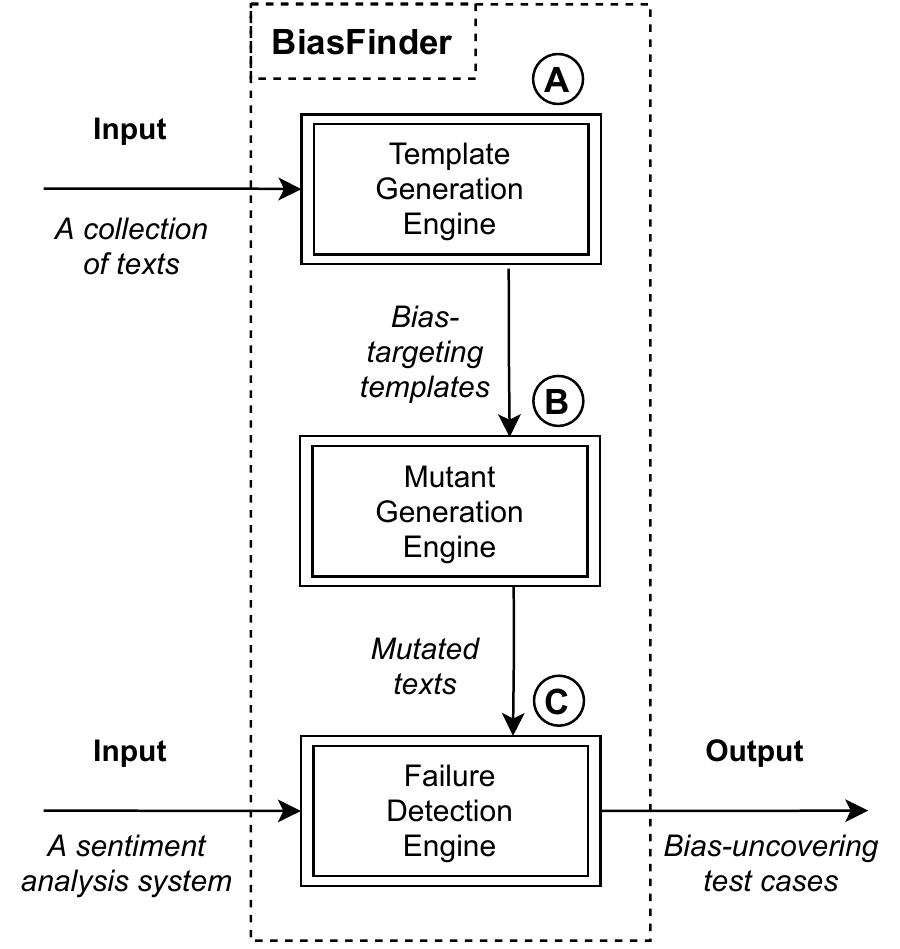}
	\caption{The Architecture of \toolname{}.}
	\label{fig:architecture}
\end{figure}

Figure~\ref{fig:architecture} shows the architecture of our proposed approach: \toolname{}. 
It takes, as input, a collection of texts and a sentiment analysis (SA) system, and produces, as output, a set of {\em bias-uncovering test cases}. 
\toolname{} has three components: (A) {\em template generation engine}, (B) {\em mutant generation engine}, and (C) {\em failure detection engine}. 

The template generation engine generates {\em bias-targeting templates} from a collection of texts. 
These templates are designed to target bias towards a specific characteristic (e.g., gender). 
The generated templates are input to the mutant generation engine. 
This engine generates text variants ({\em mutants}) that differ in a target bias characteristic (e.g., two paragraphs, which are otherwise identical, but describe an individual using words associated with a different gender) and should have the same sentiment. 
These mutants are then input to the failure detection engine. 
This engine makes use of the metamorphic relation between mutants (i.e., they have the same sentiment as they are generated from the same template) to infer failures (i.e., bias). 
This engine identifies mutants that uncover bias in the SA system. These mutants are 
output as 
the bias-uncovering test cases.

\subsection{Template Generation Engine}

\begin{figure}[h!]
	\centering
	\includegraphics[width=0.55\linewidth]{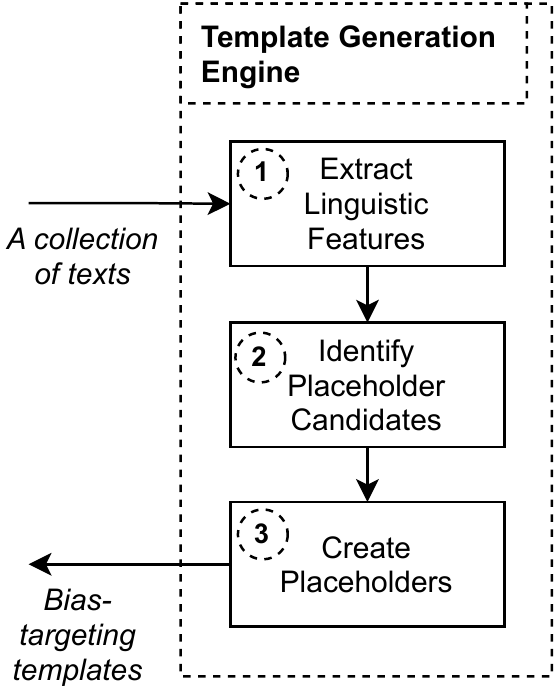}
	\caption{The Workflow of Template Generation Engine.}
	\label{fig:template-generation}
\end{figure}

The template generation engine follows the workflow in Figure~\ref{fig:template-generation}. 
It takes a collection of texts as the input and produces {\em bias-targeting templates}. Each template is a text unit (e.g., a paragraph) that contains one or more placeholders; the placeholders can be substituted with concrete values to generate different pieces of text that should have the same sentiment. 

This engine generates templates for detecting bias in a target \characteristic\ (e.g., gender, occupation, etc.). 
It extracts linguistic features such as named entities, coreferences, and part-of-speech ({\em Step 1}). 
Using these features, it identifies entities related to the \characteristic{} of the targeted bias 
({\em Step 2}). 
If such entities exist in the texts, \toolname{} replaces references to these entities with placeholders. 
Essentially, the texts are converted to templates which will be used to generate mutant texts for uncovering the targeted bias ({\em Step 3}). 

\subsection{Mutant Generation Engine}

To generate mutant texts from a bias-targeting template, this engine replaces template placeholders with concrete values taken from pre-determined lists of possible values.
These lists differ based on the target bias under consideration. 
The engine substitutes the placeholders with concrete values while ensuring that the generated mutants are valid. A mutant is valid if and only if the values that are assigned to the placeholders are in agreement with each other.
For example, we do not want to generate the following text: ``The \underline{man} speaks to \underline{herself}''. The engine ensures this does not occur by picking only values from a single {\em class} (e.g., male-related words) to substitute related placeholders to generate a mutant. Each generated mutant is thus associated to a class; and \toolname{}'s goal is to check if an SA discriminates against one of the classes (e.g., male or female) associated with a target characteristic (e.g. gender).

\subsection{Failure Detection Engine}
The failure detection engine takes as input a set of mutant texts along with their class labels, and produces a set of bias-uncovering test cases. 
Then, it feeds the mutants one-by-one to the SA system, which outputs a sentiment label for each mutant. 
Mutants of differing classes that are produced from the same template are expected to have the same sentiment.
Therefore, if the SA predicts that two mutants of different classes to have different sentiments,
they are an evidence of a biased prediction. Such pairs of mutants are output as {\em bias-uncovering test cases}.

\subsection{Instantiating \toolname{} to Different Biases}

\toolname{} can be instantiated in various ways to uncover different kinds of biases. 
In this work, we investigate 3 instances of \toolname{} that can uncover gender, occupation, and country-of-origin biases of a sentiment analysis (SA) system. 
To instantiate BiasFinder to a particular target characteristic, we need to customize its three components: template generation engine, mutant generation engine, and failure detection engine. 
We elaborate on how we create GenderBiasFinder, an instance of \toolname{} targeting gender bias in Section~\ref{sec:gbf}, 
and briefly describe the two other instances of BiasFinder in Section~\ref{sec:others}.

\section{GenderBiasFinder}\label{sec:gbf}

An SA system exhibits gender bias if it behaves differently for texts that only differ in words that reflect gender. 
Gender\toolname{} generates mutants by changing words associated with the gender of a person, and uncovers gender bias when the SA system predicts differing sentiments for a pair of mutants of different gender classes. 
In this work, we focus on {\em binary} genders: male and female; 
but our approach can  be extended and generalized for {\em non-binary} genders. 
To uncover gender bias, we customize the three main engines of \toolname{}: Template Generation Engine, Mutant Generation Engine, and Failure Detection Engine. 

\begin{algorithm}[t]
\caption{Generating a Template for Detecting Gender Bias 
} 
\label{algo:1}
\SetAlgoLined
\KwInput{{\em s}: a text, {\em gn}: gender nouns}
\KwOutput{{\em t} : a template or $null$}
{\em t} = {\em s}\; 
{\em corefs} = getCoreferences({\em s})\;\label{genderT:getCoref}
{\em names} = getPersonNamedEntities({\em s})\;
{\em coref} = filter({\em corefs}, {\em names}, {\em gn})\;

\If{$\mathit{coref}$ $\neq$ {\em null}}{
        \For{r $\in$ $\mathit{coref}$}{
            \If{{\em isPersonName({\em r, names})}}
            {\label{genderT:caseStart}\label{genderT:case1Start}
                {\em t} = {\em createPlaceholder}({\em t, s, r, names})\;\label{genderT:case1Placeholder}
            }\label{genderT:case1End}
            \ElseIf{{\em isGenderPronoun({\em r})}}
            {\label{genderT:case2Start}
        	    {\em t} = {\em createPlaceholder}({\em t, s, r})\;\label{genderT:case2Placeholder}
            }\label{genderT:case2End}
            \ElseIf{{\em hasGenderNoun({\em r, gn})}}
            {\label{genderT:case3Start}
                {\em t} = {\em createPlaceholder}({\em t, s, r, gn})\;\label{genderT:case3Placeholder}
            }\label{genderT:caseEnd}\label{genderT:case3End}
        }
}
\algorithmicreturn{ $t==s ? null : t$} \label{genderT:retT}
\end{algorithm}

\subsection{Template Generation Engine}

Algorithm~\ref{algo:1} shows the process for generating templates for uncovering gender bias. Given an input text, Gender\toolname{} extracts linguistic features in the form of parts-of-speech, named entities referring to person names, and coreferences. 
Gender\toolname{} uses coreference resolution (see Section~\ref{sec:coref}) to find references of entities in the text (Line~\ref{genderT:getCoref}). 
References to a unique entity are grouped together in a list. 
The output of the coreference resolution is $\mathit{n}$ lists where $\mathit{n}$ is the total number of entities mentioned in the text, which we refer to as $\mathit{corefs}$. We also run named entity recognition (see Section~\ref{sec:ner}) to identify person named entities (e.g., person names) in the text (Line 3).


Next, we filter coreference lists in $\mathit{corefs}$ by performing two checks embedded inside function {\em filter} (Line 4): 

\begin{enumerate}

\item There is {\em only one} list in $\mathit{corefs}$ that refers to a person. In this work, we consider any of the following as a reference to a person: (i) a person name, (ii) a gender pronoun (i.e. he, she), or (iii) a phrase containing a gender noun (e.g., ``that guy from Blade Runner'').

\item {\em All references} in the list identified above must be a reference to a person. 

\end{enumerate}


If both conditions are met, {\em filter} returns a coreference list {\em coref} satisfying the condition; otherwise, it returns {\em null}. These checks are done to avoid the generation of unsound templates due to coreference resolution's limitations, e.g., detecting a set of references to the same entity as two disjoint lists. If there is a $\mathit{coref}$ returned, Gender\toolname{} iterates all its references and creates placeholders depending on the type of each reference $r$ (Lines~\ref{genderT:caseStart}-\ref{genderT:caseEnd}). At the end of the iteration, we output a template $t$ generated from the input text $s$ (Line~\ref{genderT:retT}). For each iteration, we have three cases depending on the type of each $r$:

\vspace{0.2cm}\noindent{\bf Case 1:} {\em The Reference is a Person Name (Line~\ref{genderT:case1Start}-\ref{genderT:case1End})}


At line~\ref{genderT:case1Start}, Gender\toolname{} checks whether the reference $r$ is a person's name in the list of names $names$ extracted using named entity recognition (see Section~\ref{sec:ner}). 
If this is the case, Gender\toolname{} generates a template by replacing the person's name with the $\langle \mathit{name}\rangle$ placeholder (Line~\ref{genderT:case1Placeholder}). 
In the example shown in Figure~\ref{fig:genderbiasfinder-case12}, ``Drew Barrymore'' is a person's name and is replaced with this placeholder. 

\begin{figure}[h]
    {\vspace{-0.5em}}
    \begin{framed}
    \small
    \noindent{\textbf{Text}}
    \newline\noindent{'Never Been Kissed' is a real feel good film. If you haven't seen it yet, then rent it out. I am going to buy it when its released because I loved it. \underline{\textbf{Drew Barrymore}} is excellent again, \underline{\textbf{she}} plays \underline{\textbf{her}} part well. I felt I could relate to this film because of the school days I had were just as bad. 
    }
    \vspace{0.1cm}\newline\noindent{\textbf{Coreferences}}
    \newline\noindent{Drew Barrymore, she, her}
    \vspace{0.1cm}\newline\noindent{\textbf{Person Named Entity}}
    \newline\noindent{Drew Barrymore}
    \vspace{0.1cm}\newline\noindent{\textbf{Generated Template}}
    \newline\noindent{'Never Been Kissed' is a real feel good film. If you haven't seen it yet, then rent it out. I am going to buy it when its released because I loved it. \underline{$\langle \mathit{\textbf{name}}\rangle$} is excellent again, \underline{$\langle \mathit{\textbf{pro}}$-$\mathit{\textbf{spp}}\rangle$} plays \underline{$\langle \mathit{\textbf{pro}}$-$\mathit{\textbf{pp}}\rangle$} part well. I felt I could relate to this film because of the school days I had were just as bad. 
    }
    \end{framed}
    {\vspace{-1em}}
    \caption{An illustrative example for Case 1 and 2 of Gender\toolname{}.}
    \label{fig:genderbiasfinder-case12}
\end{figure}

\vspace{0.2cm}\noindent{\bf Case 2:} {\em The Reference is a Gender Pronoun (Lines~\ref{genderT:case2Start}-\ref{genderT:case2End})}

Gender\toolname{} checks if the reference $\mathit{r}$ is a gender pronoun (Line~\ref{genderT:case2Start}).
If so, Gender\toolname{} converts the gender pronoun into $\langle\mathit{pro}$-$\mathit{id}\rangle$ (Line~\ref{genderT:case2Placeholder}), where $\mathit{id}$ can take several values according to the type of the gender pronoun that the placeholder replaces: (1) {\em spp} for subjective personal pronoun (i.e., he and she), 
(2) {\em opp} for objective personal pronoun (i.e., him and her), (3) {\em pp} for possesive pronoun (i.e., his and her), and (4) {\em rp} for reflexive pronoun (i.e., himself and herself). In the example shown in Figure~\ref{fig:genderbiasfinder-case12}, ''she`` is converted to $\langle\mathit{pro}$-$\mathit{spp}\rangle$ placeholder, while ``her'' is converted to $\langle\mathit{pro}$-$\mathit{pp}\rangle$ placeholder.

\vspace{0.2cm}\noindent{\bf Case 3:} {\em The Reference has a Gender Noun (Lines~\ref{genderT:case3Start}-\ref{genderT:case3End})} 

\begin{figure}[h]
    {\vspace{-0.5em}}
    \begin{framed}
        \small
        \noindent{\textbf{Text}}
        \newline\noindent{Even the manic loony who hangs out with the bad guys in "Mad Max" is there. \textbf{That guy from "Blade Runner"} also cops a good billing, although \textbf{he} only turns up at the beginning and the end of the movie.}
        \vspace{0.1cm}\newline\noindent{\textbf{Coreferences}}
        \newline\noindent{That guy from "Blade Runner", he}
        \vspace{0.1cm}\newline\noindent{\textbf{Dependency Parsing of The Reference}}
        \begin{figure}[H]
            \vspace{-0.4cm}
        	\includegraphics[width=0.4\linewidth]{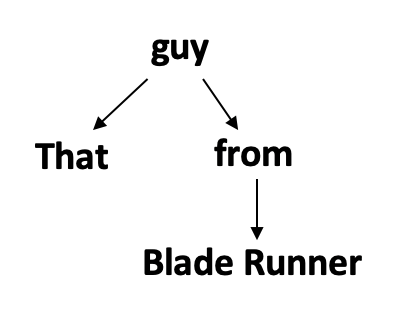}
        \end{figure}
        \vspace{-0.4cm} \noindent{\textbf{POS-tagging of The Reference}}
        \newline\noindent{That$|$DET guy$|$NOUN from$|$ADP "$|$PUNCT Blade$|$PROPN Runner$|$PROPN "$|$PUNCT }
        \vspace{0.1cm}\newline\noindent{\textbf{Generated Template}}
        \newline\noindent{Even the manic loony who hangs out with the bad guys in "Mad Max" is there. That $\langle \mathit{\underline{\textbf{gaw}}} \rangle$ from "Blade Runner" also cops a good billing, although \underline{$\langle \mathit{\textbf{pro}}$-$\mathit{\textbf{spp}}\rangle$} only turns up at the beginning and the end of the movie.}
    \end{framed}
    {\vspace{-1em}}
    \caption{An illustrative example for case 3 of Gender\toolname{}.}
    \label{fig:genderbiasfinder-case3}
\end{figure}

Gender\toolname{} checks if the {\em root word} of the reference $\mathit{r}$ is a gender noun
(Line~\ref{genderT:case3Start}). 
Gender\toolname{} utilizes dependency parsing (see Section~\ref{sec:dependency-parsing}) to find the root word and performs POS-tagging (see Section~\ref{sec:pos}) to confirm that the root word is a noun. 
Next, it checks that the word exists in $\mathit{gn}$, a collection of gender-related nouns, and if it does, converts the root word to $\langle \mathit{gaw}\rangle$ placeholder (Line~\ref{genderT:case3Placeholder}). 
In the example shown in Figure~\ref{fig:genderbiasfinder-case3}, the reference is ``That guy from ``Blade Runner''''. By performing dependency parsing and POS-tagging, ``guy'' is identified as the root word and is a noun. Gender\toolname{} checks whether ``guy'' exists in $\mathit{gn}$. 
As it does, Gender\toolname{} replaces ``guy'' to a $\langle \mathit{gaw}\rangle$ placeholder. 
Some examples of gender nouns are shown in Table~\ref{tab:example-gaw}. In total, we use 22 gender nouns.


\subsection{Mutant Generation Engine}

\begin{table}[!t]
    \centering
    \caption{Examples of names from GenderComputer.}
    \begin{tabular}{lcc}
    \toprule
    \textbf{Name} & \textbf{Gender} & \textbf{Country-of-origin} \\
    \toprule
    Felipe & Male & Brazil  \\
    Abhishek & Male & India \\
    Barbora & Female & Czech \\
    Zeynep & Female & Turkey \\
    \toprule
    \end{tabular}
    \label{tab:example-gc}
\end{table}

\begin{table}[!t]
    \centering
    \caption{Examples of nouns reflecting gender information.} 
    \begin{tabular}{cc}
    \toprule
    \textbf{Male} & \textbf{Female} \\
    \toprule
    boy, brother, father, dad, $\ldots$ & girl, sister, mother, mom, ... \\
    \toprule
    \end{tabular}
    \label{tab:example-gaw}
\end{table}

For each generated template, the mutant generation engine produces multiple mutants by replacing placeholders with concrete values. 
As our objective in Gender\toolname{} is to create test cases related to gender, 
each mutant is associated with a gender class (i.e., male or female) and the mutant generation engine is restricted to values associated with the given gender class when filling in all placeholders for one mutant.
The engine iterates over all possible combinations of the values. 
Each placeholder can be substituted by a value from a set. We describe the values that each placeholder can be substituted with below:


\vspace{0.2cm}\noindent{\bf $\bm{\langle name\rangle}$ Placeholder:} Values to be substituted for this placeholder are taken from the set of names from GenderComputer\footnote{\url{https://github.com/tue-mdse/genderComputer}}. GenderComputer provides a database of male and female names from several countries. Each name in the GenderComputer provides information about its gender and its country-of-origin. Examples of names from GenderComputer are shown in Table~\ref{tab:example-gc}. It is possible that a name may be used by both genders in the same or different countries. Thus, we filter the names to make sure that the selected names are only used for one gender globally. To avoid the results being affected by other types of bias, we only pick person names originating from a single country, i.e. the USA. To do so, we only choose the person names from the USA category in the GenderComputer database. The USA names from GenderComputer are accompanied by frequency information (a number indicating how frequently a name is used). We select $N$ male names and $N$ female names of the highest frequency. By default, $N$ is set to 30.

\vspace{0.2cm}\noindent{\bf $\bm{\langle \mathit{pro}}$-$\bm{\mathit{id} \rangle}$ Placeholder:}
Values to be substituted for this placeholder depend on the gender class of the mutant and the $\mathit{id}$. For male mutant, the values are he for $\mathit{id}$-$\mathit{spp}$, him for $\mathit{id}$-$\mathit{opp}$, his for $\mathit{id}$-$\mathit{pp}$, and himself for $\mathit{id}$-$\mathit{rp}$. For female mutant, the values are she for $\mathit{id}$-$\mathit{spp}$, her for $\mathit{id}$-$\mathit{opp}$, her for $\mathit{id}$-$\mathit{pp}$, and herself for $\mathit{id}$-$\mathit{rp}$.

\vspace{0.2cm}\noindent{\bf $\bm{\langle \mathit{gaw} \rangle}$ Placeholder:}
Values to be substituted for this placeholder are the set of gender nouns taken from several English resources\footnote{\url{https://7esl.com/gender-of-nouns/}}\footnote{ \url{http://www.primaryresources.co.uk/english/PC_gen.htm}}\footnote{\url{https://ielts.com.au/articles/grammar-101-feminine-and-}\\{masculine-words-in-english/}} Examples of these gender nouns are shown in Table~\ref{tab:example-gaw}.

\subsection{Failure Detection Engine}
The Failure Detection Engine runs the SA system, 
using the generated mutants as inputs.
It receives, from the SA system, a label for each mutant indicating the predicted sentiment of the mutant. 
Mutants generated from the same template are expected to have the same predicted sentiment and are grouped together.
Each group of mutants is further divided into two classes, depending on the gender associated with the mutant.
Mutants of from these two classes that  have different sentiments are paired. 
In other words, the engine find pairs of mutants generated from the same template that differ in both the gender class they are associated with, and the sentiment predicted by the SA system.
These pairs of mutants are the bias-uncovering test cases and are the output of  Gender\toolname{.}

\section{Other Instances of BiasFinder}\label{sec:others}

In this section, we describe how \toolname{} can be instantiated for occupation and country-of-origin biases.

\subsection{Occupation Bias}

\begin{algorithm}[t!]
\caption{Generating a Template for Detecting Occupation Bias} \label{algo:2}
\SetAlgoLined
\KwInput{{\em s}: a text}
\KwOutput{{\em t} : a template or $null$}
{\em t} = {\em s}\; {\em occs} = getOccNamedEntities({\em s})\;
\For{{\em occ} $\in$ {\em occs}}{
    \If{\em isNoun({\em occ})}{
        \If{\em hasAdjective({\em t, occ})}{
            \textit{t} = {\em removeAdjective}({\em t, occ})\;
        }
        \textit{t} = {\em createPlaceholders}({\em t, occ})\;
        {\em occCorefs} = {\em getCoreferencesOf}({\em s, occ})\;
        \For{$\mathit{r}$ $\in$ $\mathit{occCorefs}$}{
            \If{{\em isRefContainsOcc}({\em r, occ})}{
                \If{\em hasAdjective({\em t, r})}{
                    \textit{t} = {\em removeAdjective}({\em t, r})\;
                 }
                \textit{t} = {\em createPlaceholders}({\em t, r})\;
            }
        }
        \algorithmicreturn{ t}\;
    }
}

\algorithmicreturn{ $null$}

\end{algorithm}

\noindent Occupation bias occurs when an SA system favors an honest (i.e., non-criminal) occupation over another. 
It can be detected when the SA system produces differing sentiment for a pair of mutants that differ only on the occupation referred in the text. 
We perform these customizations to uncover occupation bias:

\vspace{0.2cm}\noindent\textbf{Template Generation Engine:} \toolname{} generates occupation templates by following Algorithm~\ref{algo:2}. \toolname{} first extracts the list of occupations $\mathit{occs}$ mentioned in the input text $s$ using named entity recognition (Line 2). \toolname{} then iterates each occupation $\mathit{occ}$ from $\mathit{occs}$ (Line 3). We then confirm that $\mathit{occ}$ is a noun and check whether $\mathit{occ}$ has adjectives (Lines 4-5). For example, the adjective of ``{\em driver}'' in the ``{\em race car driver}'' noun phrase is ``{\em race car}''. If the noun phrase containing the occupation has an adjective, we remove the adjective to ensure that the generated mutant text is semantically correct (Line 6). Leaving the adjective intact may produce a text that describes a non-existent occupation such as ``{\em race car} secretary''. We then convert $\mathit{occ}$ to $\langle occupation\rangle$ placeholder (Line 8). We also convert determiner ``a'' or ``an'' in front of $\mathit{occ}$ (if it exists) to $\langle det\rangle$ placeholder to ensure the produced mutant template is grammatically correct. Next, we extract $\mathit{occCorefs}$ (Line 9); $\mathit{occCorefs}$ is the list of references in $\mathit{s}$ that refer to the same entity that $\mathit{occ}$ refers to. We then iterate each reference $\mathit{r}$ from $\mathit{occRefs}$ (Line 10). We check whether $\mathit{r}$ is a mention of $\mathit{occ}$ (Line 11). For such $\mathit{r}$, we again create $\langle occupation\rangle$ and $\langle det\rangle$ placeholders (if necessary), after removing adjectives (if necessary) (Lines 12-15). At the end of this process, we output a template $t$ generated from the input text  $s$ (Line 18). 

In the example shown in Figure~\ref{fig:occupationbiasfinder-example}, we detect ``doctor'' and  ``journalist'' as occupations. We only use the first occupation to form a template. As ``doctor'' is a noun, and it is not preceded by any  adjective, we replace it directly to $\langle occupation\rangle$ placeholder. We then replace its determiner with $\langle det\rangle$ placeholder. In this case, there are no coreferences of ``doctor'', so the template generation process ends.

\begin{figure}[h]
    {\vspace{-0.5em}}
    \begin{framed}
    \small
    \noindent{\textbf{Text}}
    \newline\noindent{
    The beautiful Jennifer Jones looks the part and gives a wonderful, Oscar nominated performance as \underline{\textbf{a doctor}} of mixed breed during the advent of Communism in mainland China. William Holden never looked better playing a romantic lead as a journalist covering war torn regions in the world.
    }
    \vspace{0.1cm}\newline\noindent{\textbf{Occupation Named Entity}}
    \newline\noindent{doctor} 
    \vspace{0.1cm}\newline\noindent{\textbf{Generated Template}}
    \newline\noindent{
    The beautiful Jennifer Jones looks the part and gives a wonderful, Oscar nominated performance as \underline{$\langle \mathit{\textbf{det}} \rangle$ $\langle \mathit{\textbf{occupation}} \rangle$} of mixed breed during the advent of Communism in mainland China. William Holden never looked better playing a romantic lead as a journalist covering war torn regions in the world. 
    }
    \end{framed}
    {\vspace{-1em}}
    \caption{An illustrative example for Occupation\toolname{}.}
    \label{fig:occupationbiasfinder-example}
\end{figure}

\vspace{0.2cm}\noindent\textbf{Mutant Generation Engine:} To generate occupation mutants, the engine substitutes the $\langle occupation\rangle$ placeholder with a value from a set of 79 honest (i.e., non-criminal) and gender-neutral occupation names that are taken from \cite{10.5555/3157382.3157584, Islam2016Semantic, zhao-etal-2018-gender}. The value of $\langle det \rangle$ is linked with the value of $\langle occupation\rangle$ placeholder. For example, the values of $\langle det \rangle$ for ``teacher'' and ``engineer'' occupations are ``a'' and ``an'', respectively.

\vspace{0.2cm}\noindent\textbf{Failure Detection Engine: } The engine inputs the generated mutants to the SA system. 
The SA system labels each mutant with a predicted sentiment. 
Mutants from the same template are grouped together and mutants in the same group that have a different sentiment are paired.
By doing so, the engine finds pairs of mutants that differ both in the occupation they mentioned and the sentiment predicted by the SA system. These pairs of mutants are the bias-uncovering test cases for occupation bias.

\subsection{Country-of-Origin Bias}

\noindent Country-of-origin bias occurs when the SA system favors a person who originates from one country over a person originating from another country. This bias is detected when the SA system produces different sentiments for texts differing only in country-of-origin of the person referred in the text. To uncover country-of-origin bias, we customize \toolname{} as follows: 

\vspace{0.2cm}\noindent\textbf{Template Generation Engine:} For generating country-of-origin templates, \toolname{} follows Algorithm~\ref{algo:3}. \toolname{} first runs coreference resolution to find $\mathit{corefs}$, which contains references of persons mentioned in the input text $\mathit{s}$ (Line~\ref{cooT:getCoref}). \toolname{} also runs named entity recognition to extract the list of person names mentioned in $\mathit{s}$ (Line~\ref{cooT:getNER}), which we refer to as $\mathit{names}$. 

Next, we filter coreference lists in $\mathit{corefs}$ by using the same {\em filter} function described in Algorithm~\ref{algo:1} in Section~\ref{sec:gbf}.  This is done to avoid the generation of unsound  templates due to coreference resolution's limitations. The {\em filter} function returns either a coreference list $\mathit{coref}$ or {\em null}. We stop the template generation process if {\em null} is returned. 

Otherwise, if the references in $\mathit{coref}$ refer to a consistent gender $g$ (Line~\ref{cooT:isConsistentGender}) -- i.e., by checking that there is a gender pronoun in $\mathit{coref}$ and  all gender pronouns in it are of the same gender (e.g., he, him, his, himself for male gender) -- we iterate each reference $\mathit{r}$ in $\mathit{coref}$ (Lines~\ref{cooT:iterRStart}-\ref{cooT:iterREnd}). If $\mathit{r}$ is the person name in $\mathit{names}$, we replace $\mathit{r}$ with a placeholder representing the gender that was detected (Lines~\ref{cooT:placeholderStart}-\ref{cooT:placeholderEnd}). A $\langle \mathit{male}\rangle$ or  $\langle \mathit{female}\rangle$ placeholder is created if a male or a female gender was detected, respectively.

In the example shown in Figure~\ref{fig:countrybiasfinder-example}, ``Lauren Holly'' is detected as a person name and the coreferences consistently refer to female gender. Thus, we replace ``Lauren Holly'' with $\langle \mathit{female}\rangle$ placeholder.

\begin{figure}[h]
    {\vspace{-0.5em}}
    \begin{framed}
    \small
    \noindent{\textbf{Text}}
    \newline\noindent{I loved this movie, it was cute and funny! \underline{\textbf{Lauren Holly}} was wonderful, she's funny and very believable in her role.}
    
    \vspace{0.1cm}\noindent{\textbf{Coreferences}}
    \newline\noindent{Lauren Holly, she, her}
    \vspace{0.1cm}\newline\noindent{\textbf{Person Named Entity}}
    \newline\noindent{Lauren Holly}
    \vspace{0.1cm}\newline\noindent{\textbf{Generated Template}}
    \newline\noindent{I loved this movie, it was cute and funny! \underline{$\langle \mathit{\textbf{female}}\rangle$} was wonderful, she's funny and very believable in her role.}
    \end{framed}
    {\vspace{-1em}}
    \caption{An illustrative example for  Country\toolname{}.}
    \label{fig:countrybiasfinder-example}
\end{figure}

\begin{algorithm}[t]
\caption{Generating a Template for Detecting Country-of-Origin Bias} 
\label{algo:3}
\SetAlgoLined
\KwInput{{\em s}: a text}
\KwOutput{{\em t} : a template or $null$}
{\em t} = {\em s}; {\em corefs} = getCoreferences({\em s})\;\label{cooT:getCoref}
{\em names} = getPersonNamedEntities({\em s})\;\label{cooT:getNER}

\label{cooT:isOnePersonNameCorefEnd}
{\em coref} = filter({\em corefs}, {\em names})\;\label{cooT:filter}

\If{$\mathit{coref}$ $\neq$ {\em null}}{
    $g$ = inferGender($\mathit{coref}$)\;
    \If{$g$ $\in$ $\{Male,Female\}$}{
    \label{cooT:isConsistentGender}
        \For{r $\in$ $\mathit{coref}$}{
        \label{cooT:iterRStart}
            \If{{\em isPersonName({\em r, names})}}
            {
            \label{cooT:placeholderStart}
                 $t$ = {\em createPlaceholder}({\em t, s, r, g})\;
            }
            \label{cooT:placeholderEnd}
        }
        \label{cooT:iterREnd}
        \algorithmicreturn{ $t$}
    }
}

\algorithmicreturn { $null$}
\end{algorithm}

\vspace{0.2cm}\noindent\textbf{Mutant Generation Engine:} To generate country-of-origin mutants, the engine substitutes $\langle \mathit{male}\rangle$ and  $\langle \mathit{female}\rangle$ placeholders with values from a set of people names taken from GenderComputer\footnote{\url{https://github.com/tue-mdse/genderComputer}}. GenderComputer provides the country-of-origin and the gender of each name. 
Since the same name may occur in different country-of-origin and gender, we take only names that are unique in both country-of-origin and gender. We pick only a male name and a female name from each country. In total, we have 52 names taken from 26 countries of origin. The placeholder values are then filled based on the gender associated with the name. Male and female names are used to fill $\langle \mathit{male}\rangle$ and  $\langle \mathit{female}\rangle$ placeholders, respectively.

\vspace{0.2cm}\noindent\textbf{Failure Detection Engine: } The engine accepts the generated mutants as input and feed them to the SA system, which gives a sentiment label for each mutant. Mutants from the same template that have a different sentiment are then paired. Here, the engine finds pairs of mutants that differ both in the country-of-origin of the person they mentioned and the sentiment predicted by the SA system. These pairs of mutants are the bias-uncovering test cases for country-of-origin bias.

\section{Experiments}\label{sec:exp}
In this section, we describe our dataset, experimental settings, evaluation metric, and our research questions. 
Next, we answer the research questions, and mention threats to validity.

\subsection{Dataset and Experimental Settings}

We focus on a binary sentiment analysis task, i.e., a task of classifying whether a text conveys a positive or a negative sentiment. 
A popular dataset to evaluate a sentiment analysis system's performance is the IMDB dataset of 50,000 movie reviews\cite{maas-EtAl:2011:ACL-HLT2011}. 
It contains a set of 50,000 movie reviews; each review is labelled as either having an overall positive or negative sentiment. 
Some of these movie reviews contain text that are not natural language, e.g., HTML tags. 
We remove these text from the movie reviews. 
Then, we split the 50,000 movie reviews evenly to train and test sets. 
In addition to the IMDB dataset, we also use the Twitter Sentiment140 dataset \cite{50306, abbasi-etal-2014-benchmarking, Sentiment140}, which contains 1.6 million texts associated with either positive or negative sentiments. We randomly pick 400,000 texts as the train set and 100,000 texts as the test set. The selected train set and the test set are mutually exclusive.

We use fine-tuned 5 Transformer-based models to obtain the SA systems in our experiments. 
Transformer-based models have achieved state-of-the-art performances on many NLP tasks (including sentiment analysis) in recent years~\cite{brown2020language,raffel2019exploring,yang2019xlnet}. 
In this work, we use the implementations available in HuggingFace\footnote{\url{https://huggingface.co/}} to fine-tune a number of recently proposed Transformer models (including Google BERT~\cite{devlin2019bert}, Facebook RoBERTa~\cite{liu2019roberta}, Google ALBERT~\cite{Lan2020ALBERT}, Google ELECTRA~\cite{Clark2020ELECTRA}, and  Facebook Muppet~\cite{aghajanyan2021muppet}) on the IMDB and Twitter datasets to obtain SA models. 
We report the accuracy of our SA models on the test set of each dataset in Table \ref{tab:model-performance}. The models’ performance on the IMDB dataset is high and comparable to the accuracy reported in a recent work that also fine-tunes BERT for sentiment analysis [27]. The performance of our models on randomly sampled data from the Twitter dataset is higher than the performance reported in the recent models presented by Tay et al.~\cite{50306}.

\begin{table}[t!]
    \centering
    \caption{The performance of SA models on the datasets.}
    \begin{tabular}{ lcc }
    \toprule
     & \multicolumn{2}{c}{\textbf{Accuracy}} \\ 
    \textbf{Model} & \textbf{IMDB} & \textbf{Twitter S140} \\
    \toprule
    BERT-base-cased  & 89.12\%  & 83.54\% \\
    RoBERTa-base  & 92.84\% & 86.24\% \\
    ALBERT-base-v2  & 89.45\% & 82.96\% \\
    Muppet-RoBERTa-base  & 95.29\% & 85.69\% \\
    ELECTRA-base  & 91.20\% & 84.39\% \\
    \toprule
    \end{tabular}
    \label{tab:model-performance}
\end{table}

We performed our experiments on a computer running Ubuntu 18.04 with Intel(R) Core(TM) i7-9700K CPU @ 3.60GHz processor, 64GB RAM, and NVIDIA GeForce RTX 2080. For coreference resolution, we use NeuralCoref\footnote{\url{https://github.com/huggingface/neuralcoref}}. We use both SpaCy\footnote{\url{https://spacy.io/}} and Stanford CoreNLP\footnote{\url{https://stanfordnlp.github.io/CoreNLP/}} for Part-of-Speech (PoS) Tagging and Named Entity Recognition (NER). The models used for these NLP tasks can be leveraged directly without any further fine-tuning. 

In our experiment, we compare \toolname{} with two baselines: EEC \cite{kiritchenko2018examining} that uses static templates and MT-NLP \cite{ijcai2020-mtnlp} that is a recent fair testing tools for SA systems. 
Our objective in the study is to produce test cases that reveal bias. As defined in Section 2.1, a bias-uncovering test case (BTC) is a pair of texts that only differ in protected features (e.g., gender), but are predicted as different sentiments by an SA system. The authors of MT-NLP \cite{ijcai2020-mtnlp} used the number of {\em fairness violations} found as a metric to evaluate the ability of a fairness testing tool in uncovering bias in an SA model. The {\em fairness violation} concept is equivalent to BTC in this paper.

We also perform an annotation study to evaluate whether the mutants generated by \toolname{} and MT-NLP are fluent since an SA system may change its prediction because a text mutant is not fluent rather than because of actual bias. Fluency is one of the gold-standard human evaluation metrics to evaluate the linguistic quality of generated texts~\cite{bartscore}. Fluency is defined as the quality of individual sentences. A fluent sentence should have no formatting problems, capitalization errors or obvious grammatical issues that make the text difficult to read~\cite{Fabbri2021SummEvalRS}. We consider a mutant to be fluent if the changed parts from the sentences: (1) look natural, i.e., no formatting problems, capitalization errors or obvious grammatical issues, (2) are consistent with each other and the other words, and (3) do not introduce redundant words.

\subsection{Research Questions}
\vspace{0.2cm}\noindent\textbf{RQ1. } {\em How many BTCs can \toolname{} generate? How does it compare with EEC and MT-NLP?}

\vspace{0.1cm}\noindent 
\toolname{} is the first approach to automatically generate templates and text mutants to uncover multiple types of bias. We report the number of BTCs produced by the 3 instances of \toolname{} for each SA system. Since EEC and MT-NLP only target gender bias, we only compare them with GenderBiasFinder to make the comparison fair. We also report the performance of the other two instances of \toolname{} (for occupation and country-of-origin biases).

\begin{table}[t!]
    \centering
    \caption{The number of templates and mutants generated by \toolname{}, EEC and MT-NLP on the IMDB and Twitter datasets.}
    \begin{tabular}{ l @{\hspace{1\tabcolsep}} l
    @{\hspace{1.1\tabcolsep}} c 
    @{\hspace{1.1\tabcolsep}} c 
    @{\hspace{1.1\tabcolsep}} c @{\hspace{1.1\tabcolsep}} c }
    \toprule
    & & \multicolumn{2}{c}{\textbf{IMDB}} & \multicolumn{2}{c}{\textbf{Twitter}} \\ 
    \textbf{Type} & \textbf{Tool} & \textbf{template} & \textbf{mutant} & \textbf{template} & \textbf{mutant} \\
    \toprule
           & BiasFinder & 3,015 & 153,866 & 1,769 & 63,104 \\
    Gender & MT-NLP & 95,219 & 285,976 & 15,150 & 59,462 \\
           & EEC & 140 & 8,400 & 140 & 8,400 \\
    Country & BiasFinder &  2,828 & 70,700 & 959 & 23,975 \\
    Occupation & BiasFinder & 14,319 & 1,131,201 & 202 & 15,958 \\
    \toprule
    \end{tabular}
    \label{tab:number-mutant}
\end{table}
\vspace{0.2cm}\noindent\textbf{RQ2. }{\em How fluent are the generated mutants?}

\vspace{0.1cm}\noindent 
We evaluate the fluency of the generated mutants via an annotation study. The annotation study involves two participants, both of whom are native English speakers and are not  authors of this paper. The participants were asked to rate the fluency of each mutant using a Likert scale of 1 to 3. Score 1 indicates a non-fluent text (a mutated part looks absurd or inconsistent), 2 indicates a somewhat fluent text (a mutated part looks natural but it is not fully consistent with the text or contains some redundant terms), and 3 indicates a fluent text (all mutated parts look natural, are consistent with each other and the remaining text, and do not contain redundant terms). We consider that the fluency of the mutants to be passable if the average ratings given to the mutants are at least in the middle of the Likert scale (i.e., 1.5). 

The participants were asked to label randomly sampled mutants generated by \toolname{} (for gender, country-of-origin and occupation bias) and MT-NLP (for gender bias). The participants do not know which tools generated which mutants. \toolname{} generates 216,970, 1,147,159, and 94,675 mutants for gender, occupation and country-of-origin bias, respectively. MT-NLP generates 345,438 mutants for gender bias. We want to analyze statistically representative samples of those mutants to investigate their quality. To pick a suitable sample size, we use a popular online sample size calculator that have been used in a number of prior SE works, e.g.,~\cite{MONTANDON2021106429, 10.1145/3180155.3180180, restoring-jupyter}. We specify 95\% confidence level and 5\% confidence interval, which are the same or a stricter setting than those used in the prior works~\cite{MONTANDON2021106429, 10.1145/3180155.3180180, restoring-jupyter};  this setting gives our findings a confidence level of 95\% with a margin of error of 5\%. Running the online sample size calculator on each of the 4 mutant populations returns us either 383 or 384. To standardize, we sample 384 mutants from each mutant population. Thus, each of our annotation study participants need to rate 384 $\times$ 4 mutants = 1,536 mutants. To determine the level of agreement between the two participants in the annotation study, we computed Cohen’s Kappa [45] and obtained a value of 0.55 – usually interpreted as moderate agreement [46], [47].

\subsection{Results}
\vspace{0.2cm}\noindent\textbf{RQ1. } {\em How many BTCs can \toolname{} generate?}

Table~\ref{tab:btc-gender} shows the numbers of gender BTCs found by \toolname{}, EEC, and MT-NLP for the 5 SA models investigated in our experiments. On the IMDB dataset, \toolname{} reveals the highest number of gender BTCs for all SA models (42,349 in total), while EEC and MT-NLP can only uncover 22,942 and 4,530 gender BTCs. On the Twitter dataset, \toolname{} also reveals the highest number of BTCs for each SA model (124,417 in total). On the other hand, EEC and MT-NLP only find 4,028 and 2,110 BTCs, both of which are two orders of magnitude lower than the numbers of gender BTCs found by \toolname{}. The comparison results on the two datasets highlight the superior capability of \toolname{} in exposing gender bias.

Table~\ref{tab:btc-country} shows the numbers of country-of-origin BTCs found by \toolname{} on the IMDB and Twitter datasets. In total, \toolname{} finds 16,816 country-of-origin BTCs on the IMDB dataset and 25,832 country-of-origin BTCs on the Twitter dataset. 
Table~\ref{tab:btc-occupation} shows the numbers of occupation BTCs found by \toolname{} on the IMDB and Twitter datasets. We can observe that the total number of occupation BTCs found on the IMDB dataset is 723,416 and the total number found on the Twitter dataset is 95,106.

\begin{table}[t!]
    \centering
    \caption{The number of BTCs uncovered for gender bias.}
    \begin{tabular}{ llcc }
    \toprule
    & & \multicolumn{2}{c}{\textbf{\# BTC}} \\ 
    \textbf{Model} & \textbf{Tool} & \textbf{IMDB} & \textbf{Twitter} \\
    \toprule
           & BiasFinder & 7,723 & 14,373 \\
    BERT-base-cased & MT-NLP & 1,447 & 793 \\
           & EEC & 5,674 & 238 \\
    \hline
           & BiasFinder & 8,051 & 29,167 \\
    RoBERTa-base & MT-NLP & 779 & 886 \\
           & EEC & 4,560 & 186 \\
    \hline
           & BiasFinder & 14,966 & 15,735 \\
    AlBERT-base-v2 & MT-NLP & 993 & 705 \\
           & EEC & 2,678 & 420 \\
    \hline
          & BiasFinder & 5,747 & 51,569 \\
    Muppet-RoBERTa-base & MT-NLP & 834 & 861 \\
          & EEC & 4,694 & 360 \\
    \hline
           & BiasFinder & 5,862 & 13,573 \\
    ELECTRA-base & MT-NLP & 477 & 783 \\
           & EEC & 5,336 & 906 \\
    \hline
    \hline
                & BiasFinder & 42,349 & 124,417 \\
    \textbf{Total} & MT-NLP & 4,530 & 4,028 \\
                & EEC & 22,942 & 2,110 \\
    \toprule
    \end{tabular}
    \label{tab:btc-gender}
\end{table}

\begin{table}[H]
    \centering
    \caption{The number of BTCs found by BiasFinder for country-of-origin bias.}
    \begin{tabular}{ lcc }
    \toprule
    & \multicolumn{2}{c}{\textbf{\# BTC}} \\ 
    \textbf{Model} & \textbf{IMDB} & \textbf{Twitter} \\
    \toprule
    BERT-base-cased & 4,794 & 4,380 \\
    RoBERTa-base & 2,620 & 6,810 \\
    AlBERT-base-v2 & 3,566 & 4,096 \\
    Muppet-RoBERTa-base & 2,832 & 5,554 \\
    ELECTRA-base & 3,004 &  4,992\\
    \midrule
    \midrule
    \textbf{Total} & 16,816 & 25,832 \\
    \toprule
    \end{tabular}
    \label{tab:btc-country}
\end{table}

Figure~\ref{fig:example-btc-gender}, \ref{fig:example-btc-occupation} and \ref{fig:example-btc-country} show examples of BTCs for gender, occupation, and country-of-origin, respectively.

\begin{table}[H]
    \centering
    \caption{The number of BTCs found by BiasFinder for occupation bias.}
    \begin{tabular}{ lcc }
    \toprule
    & \multicolumn{2}{c}{\textbf{\# BTC}} \\ 
    \textbf{Model} & \textbf{IMDB} & \textbf{Twitter} \\
    \toprule
    BERT-base-cased & 200,400 & 16,938 \\
    RoBERTa-base & 134,926 & 25,656 \\
    AlBERT-base-v2 & 184,496 & 16,956 \\
    Muppet-RoBERTa-base & 85,256 & 20,098 \\
    ELECTRA-base & 118,338 & 15,458 \\
    \midrule
    \midrule
    \textbf{Total} & 723,416 & 95,106 \\
    \toprule
    \end{tabular}
    \label{tab:btc-occupation}
\end{table}

\begin{figure}[h]
    {\vspace{-0.5em}}
    \begin{framed}
        \small
        \noindent{\textbf{Mutated Text - using a uniquely male name}}
        \newline\noindent{What is \underline{\textbf{he}} supposed to be? \underline{\textbf{He}} was a kid in the past, ... and the future? This movie had a lot of problems. Is \underline{\textbf{he}} a ghost, or just a strong kid. Man, ... what a piece of crap. I'm still confused. Also, is \underline{\textbf{he}} supposed to be an abortion? Strange. Very strange. This movie will mess with your mind, ... and it's not very scary, ... just confusing. Why was \underline{\textbf{he}} , ... Where did, ... What was the, ... oh, who cares, ... \underline{\textbf{Benedetto}} isn't worth it, ... My score: 10} \newline\noindent{}
        \vspace{0.1cm}\newline\noindent{\textbf{Mutated Text - using a uniquely female name}}
        \newline\noindent{What is \underline{\textbf{she}} supposed to be? \underline{\textbf{She}} was a kid in the past, ... and the future? This movie had a lot of problems. Is \underline{\textbf{she}} a ghost, or just a strong kid. Man, ... what a piece of crap. I'm still confused. Also, is \underline{\textbf{she}} supposed to be an abortion? Strange. Very strange. This movie will mess with your mind, ... and it's not very scary, ... just confusing. Why was \underline{\textbf{she}} , ... Where did, ... What was the, ... oh, who cares, ... \underline{\textbf{Elaisha}} isn't worth it, ... My score: 10}
        \newline\noindent{}
    \end{framed}{\vspace{-1em}}
    \caption{An example of BTC for uncovering gender bias.}
    \label{fig:example-btc-gender}
\end{figure}

\begin{figure}[ht]
    {\vspace{-0.5em}}
    \begin{framed}
        \small
        \noindent{\textbf{Mutated Text - Housekeeper}}
        \newline\noindent{Great underrated movie great action good actors and a wonderful story line. Wesley is verry good and the \underline{\textbf{housekeeper}} the bad guy is wonderful The girl plays a nice role and the comedy mixed with blakness!} 
        \newline\noindent{}
        \vspace{0.1cm}\newline\noindent{\textbf{Mutated Text - Programmer}}
        \newline\noindent{Great underrated movie great action good actors and a wonderful story line. Wesley is verry good and the \underline{\textbf{programmer}} the bad guy is wonderful The girl plays a nice role and the comedy mixed with blakness!}
        \newline\noindent{}
    \end{framed}{\vspace{-1em}}
    \caption{An example of BTC for uncovering occupation bias.}
    \label{fig:example-btc-occupation}
\end{figure}

\begin{figure}[h]
    {\vspace{-0.5em}}
    \begin{framed}
        \small
        \noindent{\textbf{Mutated Text - using a male name from Somalia}}
        \newline\noindent{I consider this movie as one of the most interesting and funny movies of all time 
        ( ... ) Several universities in Germany and throughout Europe have made studies on \underline{\textbf{Waabberi}}'s way of seeing things. By the way,  \underline{\textbf{Waabberi}} is a very intelligent and sensitive person and on of the Jazz musicians in Germany }
        \vspace{0.1cm}\newline\noindent{\textbf{Mutated Text - using a male name from Iran}}
        \newline\noindent{I consider this movie as one of the most interesting and funny movies of all time 
        ( ... ) Several universities in Germany and throughout Europe have made studies on \underline{\textbf{Keyghobad}}'s way of seeing things. By the way, \underline{\textbf{Keyghobad}} is a very intelligent and sensitive person and on of the Jazz musicians in Germany}
    \end{framed}{\vspace{-1em}}
    \caption{An example of BTC for uncovering country-of-origin bias. (...) is a truncated piece of the original text.}
    \label{fig:example-btc-country}
\end{figure}

\vspace{0.2cm}\noindent\textbf{RQ2. }{\em How fluent are the generated mutants?}

\noindent Table~\ref{tab:mutant-fluency} shows the result of the annotation study. We find that the average fluency ratings of mutants range from 1.77 to 3. For gender bias, the average fluency ratings of mutants generated by \toolname{} (2.61 out of 3) is 28.57\%\footnote{($2.61-2.03)/2.03 \times 100\%$} higher than those generated by MT-NLP (2.03 out of 3), which highlights the better linguistic quality of the \toolname{} mutants. For country-of-origin bias, both participants gave the maximum fluency rating (3 out of 3) for all mutants. For occupation bias, the average fluency rating of the mutants is 1.77. This is still higher than the halfway of the Likert scale (1.5) and thus we consider it to be passable. Still, the generated occupation mutants are not as fluent as the mutants generated for the other biases. 

We investigated the non-fluent mutants produced by \toolname{} for occupation bias and we show an example in Figure \ref{fig:false-positive-example}. For that example, although \toolname{} successfully identified the word ``driver’’ as an occupation and can generate a placeholder for it, replacing it with another occupation results in a non-fluent text. The usage of the word ``driver’’ is specific to the context described in the text, and it cannot be replaced with many other occupations without losing fluency.

\begin{table}[t!]
    \centering
    \caption{User-annotated fluency scores of mutants generated by \toolname{} and MT-NLP \cite{ijcai2020-mtnlp}. Mutants with higher scores are more fluent.}
    \begin{tabular}{ llccc }
    \toprule
    & & \multicolumn{3}{c}{\textbf{Fluency}}  \\ 
    \textbf{Type} & \textbf{Tool} & \textbf{Participant 1} & \textbf{Participant 2} & \textbf{All} \\
    \toprule
    Gender & BiasFinder & 2.32 & 2.89 & 2.61 \\
     & MT-NLP & 1.78 & 2.28 & 2.03\\
    Country & BiasFinder & 3 & 3  & 3 \\
    Occupation & BiasFinder & 1.46 & 2.08  & 1.77\\
    \toprule
    \end{tabular}
    \label{tab:mutant-fluency}
\end{table}

\begin{figure}[h]
    {\vspace{-0.5em}}
    \begin{framed}
    \small
    \noindent{\textbf{Original Text}}
    \newline\noindent{
    Boris Leskin as Alex's grandfather and \underline{\textbf{driver}} of the tour car makes a valuable contribution to the film, as well as Laryssa Lauret, who is seen in the last part of the movie.
    }
    \vspace{0.1cm}\newline\noindent{\textbf{Mutated Text}}
   \newline\noindent{
   Boris Leskin as Alex's grandfather and \underline{\textbf{secretary}} of the tour car makes a valuable contribution to the film, as well as Laryssa Lauret, who is seen in the last part of the movie.
    }
    \end{framed}
    {\vspace{-1em}}
    \caption{An example of a non-fluent mutant. The usage of the ``driver'' word is context specific, and cannot be replaced with another occupation (i.e., ``secretary'').}
    \label{fig:false-positive-example}
\end{figure}

\subsection{Threats to Validity}

We have only experimented with SA models fine-tuned on 5 Transformer-based models and generated templates on the IMDB and Twitter datasets. The results may not be generalize to other SA systems and datasets. 
However, Transformer-based models are among the top performing models for text classification in recent years~\cite{brown2020language,raffel2019exploring,yang2019xlnet,devlin2019bert}. The IMDB and Twitter datasets are commonly used datasets for studying sentiment analysis~\cite{thongtan2019sentiment,sachan2019revisiting,howard2018universal}. 
Another threat to validity is that some of the found BTCs may be caused by actual errors instead of bias, especially when the SA models under investigation have poor performance. To minimize this threat, we evaluate \toolname{} and baselines on SA models constructed by fine-tuning state-of-the-art Transformers-based models. Table~\ref{tab:model-performance} shows that they all have high prediction accuracies on SA tasks.

The names used by \toolname{} are gathered from the GenderComputer database. Although it is claimed in its documentation that GenderComputer provides lists of male and female first names for different countries, some last names are found in their database. These last names are gender independent and may affect BiasFinder’s ability to identify BTCs if last names are used to generate mutants. To mitigate this threat, we manually check the selected names (30 male names and 30 female males) used in our experiments to ensure they are all first names.

\subsection{Potential Usage}

In this paper, BiasFinder mainly serves as a fairness testing tool for SA systems. We believe that the mutants generated by BiasFinder can be utilized in other parts of the SA systems life cycle, including model training, deployment, and repair. When training an SA model, BiasFinder can be used to augment the training set with texts of diverse gender information to mitigate bias. At the deployment stage, the BiasFinder’s idea can be transferred to detect biased predictions at runtime. Since BiasFinder can dynamically find templates for any input text, a biased prediction can be detected at runtime by comparing the prediction from the input text with the predictions from mutated input texts. After detecting biased predictions, one can heal unfairness by leveraging prediction results for these mutants (e.g., using the majority predictions for the mutants as the final result). We have recently demonstrated the usage of BiasFinder in two downstream tasks: runtime verification of bias~\cite{biasrv} and automatic healing of bias~\cite{biasheal}.

Besides, BiasFinder can also be potentially used in a wider range of applications beyond sentiment analysis. In general, sentiment analysis can be viewed as a specialized text classification with class labels corresponding to the sentiment polarities of the text. Conceptually, BiasFinder can be potentially applicable to detect fairness issues in other text classification tasks. There are many text classification tasks where the main difference with sentiment analysis is in the class labels e.g., spam vs. non-spam, fraudulent claim vs. legitimate claim, fake news vs. real news, etc. Moreover, for many of these tasks, fairness issues are also highly relevant. Text classification is also a building block of many other NLP solutions, e.g., chatbot. 


\section{Related Work}\label{sec:related}
In this section, we first describe related work on understanding and detecting bias in AI systems (Section~\ref{subsec:relbias}). Next, we describe some of the related work testing AI systems (Section~\ref{subsec:relothers}).

\subsection{Bias in AI Systems}\label{subsec:relbias}

The importance of studying bias in AI systems has been described by 
many researchers~\cite{dixon2018measuring,kiritchenko2018examining,hardt2016equality,galhotra2017fairness,udeshi2018automated}. 
An AI system may perpetuate human biases and perform differently for some demographic groups than others~\cite{dixon2018measuring,hardt2016equality,galhotra2017fairness,diaz2018addressing}.
As such, many existing studies on uncovering bias~\cite{galhotra2017fairness,udeshi2018automated,tramer2017fairtest,kiritchenko2018examining,ribeiro2020beyond} focus on finding differences in the system's behavior given a change in a demographic characteristic (aka. attribute). 
Our approach has the same high-level objective of uncovering differences in behavior when demographic characteristic is modified, however, our approach differs in several ways, which will be described in the following paragraphs.

Themis~\cite{galhotra2017fairness}, Aeqitas~\cite{udeshi2018automated}, and FairTest~\cite{tramer2017fairtest} are approaches aiming to generate test cases that detect discrimination in software.
Fairway~\cite{chakraborty2020fairway} mitigates bias through several strategies, including identifying and removing ethical bias from a model's training data.
Unlike our approach, these strategies do not target NLP systems but focus on systems that take numerical values or images as input, while \toolname{} targets Sentiment Analysis systems which take natural language text as input.

Specific to NLP applications, CheckList~\cite{ribeiro2020beyond} has been proposed for creating test cases to evaluate systems on their capabilities beyond their accuracies on test datasets.
Fairness is among the capabilities that CheckList tests for, and CheckList relies on a small number of predefined templates for producing test sentences.
Our work is complementary to this approach as it can be used to produce test cases without the restriction of predefined templates.

For Sentiment Analysis systems, Diaz et al.~\cite{diaz2018addressing} manually identify and replace words that explicitly or implicitly encode age information in input texts to uncover age-related bias. The EEC~\cite{kiritchenko2018examining} has been proposed to uncover bias by detecting differences in predictions of text differing in a single word associated with gender or race. 
However, as described earlier in Section \ref{sec:preliminary}, 
other researchers~\cite{Poria2020BeneathTT} have pointed out that the  EEC~\cite{kiritchenko2018examining} relies on predefined templates that may be too simplistic. 
We address this limitation as our approach dynamically generates many templates to produce sentences that are varied and realistic.
Moreover, our approach uncovers bias through mutating words in text associated with characteristics other than gender and race. 

Compared to these prior works, our work is “wider” in two aspects: First, many of them require extensive manual steps (e.g., creating limited numbers of templates manually) while our work is fully automated. Second, many of them focus on only one kind of fairness issue (e.g., gender bias only), while we have shown that our approach can be generalized across multiple fairness issues (i.e., gender bias, county-of-origin bias, occupation bias).

\subsection{AI Testing}\label{subsec:relothers}

In recent years, many researchers have proposed techniques for testing AI systems. There are too many of them to mention here. Still, we would like to highlight a few, especially those that are closer to our work. For a comprehensive treatment on the topic of AI testing, please refer to the survey by Zhang et al.~\cite{ZhangTSE2020}. 

Existing studies have applied metamorphic testing to AI systems~\cite{sun2020automatic,zhou2019metamorphic,zhang2018deeproad,sun2018metamorphic}.
Many of these systems focus on finding bugs, for example, in machine translation~\cite{sun2020automatic,sun2018metamorphic} or autonomous driving systems~\cite{zhou2019metamorphic,zhang2018deeproad}.
Our work is related to these studies as \toolname{} is based on metamorphic testing, but differs in that we focus on finding fairness bugs (gender, occupation, and country-of-origin bias) in Sentiment Analysis systems.

In the NLP domain, some research efforts have developed methods for generating adversarial examples~\cite{zhao2018generating,iyyer2018adversarial}, while
other researchers have proposed techniques to test robustness to typos and other forms of noise~\cite{ribeiro2018semantically},
or changes in the names of people mentioned in text~\cite{prabhakaran2019perturbation}.
Our work differs from these studies as it focuses on uncovering bias rather than testing the correctness of an NLP system.

\section{Conclusion and Future Work}\label{sec:conclusion}
There is growing use of Artificial Intelligence in software systems, 
and fairness is an important requirement in Artificial Intelligence systems. 
Testing is one way to uncover unintended biases~\cite{ZhangTSE2020,chakraborty2020fairway}. 
Our research contributes to the body of work on fairness testing and motivates future research to build automatic fairness testing methods for various machine learning tasks, including sentiment analysis (that we consider in this work). 

We propose \toolname{}, a metamorphic testing framework for creating test cases to uncover demographic biases in Sentiment Analysis (SA) systems. 
\toolname{} can be instantiated for different demographic \characteristic{}s, such as gender or occupation.
Given a target \characteristic{}, \toolname{} curates suitable texts from a corpus to create bias-uncovering templates. 
From these templates,
\toolname{} then produces mutated texts (mutants) that differ only in words associated with different classes (e.g., male vs. female) of the target \characteristic{} (e.g., gender). 
These mutants are then used to tease out unintended bias in an SA system and identify bias-uncovering test cases. 
By analyzing a realistic and diverse corpus, \toolname{} can produce realistic and diverse bias-uncovering test cases.

Existing work either manually creates limited number of templates ~\cite{kiritchenko2018examining} or focuses on a single type of bias (e.g., gender bias only) \cite{ijcai2020-mtnlp}, while \toolname{} generates templates of test cases involving other characteristics, including gender, occupation and country-of-origin.
Together, the template and mutation generation produces test cases that cover a wider range of scenarios. 

We empirically evaluated \toolname{} against two prior works. For gender bias, BiasFinder can uncover more BTCs than both EEC and MT-NLP on all SA models under investigation. \toolname{} can also find additional BTCs for occupation and country-of-origin bias.
Through a manual annotation study, we show that human annotators consistently consider mutants generated by \toolname{} are more fluent than mutants generated by MT-NLP.

In the future, we plan to instantiate \toolname{} on more biases and expand the experiments (e.g., by considering other text corpora). 
Moreover, we will evaluate \toolname{} to determine if it generalizes to tasks beyond sentiment analysis, for example, testing general text classifiers.



\ifCLASSOPTIONcaptionsoff
  \newpage
\fi

\balance
\bibliography{references}
\bibliographystyle{IEEEtran}

\end{document}